\documentclass[preprint]{emulateapj}

\usepackage{natbib}

\slugcomment{To appear in the Astrophysical Journal}

\shorttitle{BLAST06: Calibration and Flight Performance}
\shortauthors{Truch~et al.}

\newcommand{\h}{^{\mathrm h}}
\newcommand{\m}{^{\mathrm m}}

\newcommand{\IRAS}{{\it IRAS}}

\newcommand{\Herschel}{{\it Herschel}}

\usepackage{textcomp}
\renewcommand{\micron}{\hbox{\textmu}m}

\begin{document}

\title{The Balloon-borne Large Aperture Submillimeter Telescope (BLAST) 2006: 
       Calibration and Flight Performance}

\author{
  Matthew~D.~P.~Truch,\altaffilmark{1}
  Peter~A.~R.~Ade,\altaffilmark{2}
  James~J.~Bock,\altaffilmark{3,4}
  Edward~L.~Chapin,\altaffilmark{5}
  Mark~J.~Devlin,\altaffilmark{1}
  Simon~R.~Dicker,\altaffilmark{1}
  Matthew~Griffin,\altaffilmark{2}
  Joshua~O.~Gundersen,\altaffilmark{6}
  Mark~Halpern,\altaffilmark{5}
  Peter~C.~Hargrave,\altaffilmark{2}
  David~H.~Hughes,\altaffilmark{7}
  Jeff~Klein,\altaffilmark{1}
  Gaelen~Marsden,\altaffilmark{5}
  Peter~G.~Martin,\altaffilmark{8,9}
  Philip~Mauskopf,\altaffilmark{2}
  Lorenzo~Moncelsi,\altaffilmark{2}
  C.~Barth~Netterfield,\altaffilmark{9,10}
  Luca~Olmi,\altaffilmark{11,12}
  Enzo~Pascale,\altaffilmark{2}
  Guillaume~Patanchon,\altaffilmark{13}
  Marie~Rex,\altaffilmark{1}
  Douglas~Scott,\altaffilmark{5}
  Christopher~Semisch,\altaffilmark{1}
  Nicholas~E.~Thomas,\altaffilmark{6}
  Carole~Tucker,\altaffilmark{2}
  Gregory~S.~Tucker,\altaffilmark{14}
  Marco~P.~Viero,\altaffilmark{9}
  Donald~V.~Wiebe\altaffilmark{5,10}
}

\altaffiltext{1}{Department of Physics and Astronomy, University of Pennsylvania, 209 South 33rd Street,
                 Philadelphia, PA 19104; {\url{matthew@truch.net}}}
\altaffiltext{2}{Department of Physics \& Astronomy, Cardiff University, 5 The Parade, Cardiff, CF24~3AA, UK}
\altaffiltext{3}{Jet Propulsion Laboratory, Pasadena, CA 91109-8099}
\altaffiltext{4}{Observational Cosmology, MS 59-33, California Institute of Technology, Pasadena, CA 91125}
\altaffiltext{5}{Department of Physics \& Astronomy, University of British
                 Columbia, 6224 Agricultural Road, Vancouver, BC V6T~1Z1,
                 Canada}
\altaffiltext{6}{Department of Physics, University of Miami, 1320 Campo Sano Drive, Coral Gables, FL 33146}
\altaffiltext{7}{Instituto Nacional de Astrof\'{\i}sica \'Optica y Electr\'onica (INAOE),
                  Aptdo. Postal 51 y 72000 Puebla, Mexico}
\altaffiltext{8}{Canadian Institute for Theoretical Astrophysics, University of Toronto,
                 60 St. George Street, Toronto, ON M5S~3H8, Canada}
\altaffiltext{9}{Department of Astronomy \& Astrophysics, University of Toronto, 50 St. George Street,
                  Toronto, ON  M5S~3H4, Canada}
\altaffiltext{10}{Department of Physics, University of Toronto, 60 St. George Street, Toronto, ON M5S~1A7, Canada}
\altaffiltext{11}{University of Puerto Rico, Rio Piedras Campus, Physics Dept., Box 23343, UPR station,
                  San Juan, Puerto Rico}
\altaffiltext{12}{INAF - Osservatorio Astrofisico di Arcetri, Largo E. Fermi 5, I-50125 Firenze, Italy}
\altaffiltext{13}{Universit{\'e} Paris Diderot, Laboratoire APC, 10 rue Alice Domon et L{\'e}onie Duquet 75205 Paris, France}
\altaffiltext{14}{Department of Physics, Brown University, 182 Hope Street, Providence, RI 02912}

\begin{abstract}
The Balloon-borne Large Aperture Submillimeter Telescope (BLAST) operated
successfully during a 250-hour flight over Antarctica in December 2006
(BLAST06)\@.  As part of the calibration and pointing procedures, the red
hypergiant star VY\,CMa was observed and used as the primary calibrator.
Details of the overall BLAST06 calibration procedure are discussed.  The
1-$\sigma$ uncertainty on the absolute calibration is accurate to 9.5, 8.7, and
9.2\% at the 250, 350, and 500\,\micron\ bands, respectively.  The errors are
highly correlated between bands resulting in much lower error for the derived
shape of the 250--500\,\micron\ continuum.  The overall pointing error is
$<$\,5\arcsec\ rms for the 36, 42, and 60\arcsec\ beams.  The performance of
the optics and pointing systems is discussed.
\end{abstract}


\section{Introduction}     \label{sec:intro}
The December 2006 flight of the Balloon-borne Large Aperture Submillimeter
Telescope (BLAST) incorporated a 1.8-m parabolic primary mirror and
large-format bolometer arrays operating at 250, 350, and 500\,\micron.  A
complete description of the BLAST instrument can be found in
\citet{pascale2008}.  The BLAST bands sample the peak of the spectral energy
distribution (SED) for cool dust (${\sim}\,10$\,K).  Astronomical signals at
these wavelengths are difficult or impossible to access from even the best
ground-based sites.  As a result, BLAST has the ability to conduct unique
Galactic and extragalactic submillimeter surveys with sub-arcminute resolution
and high sensitivity.  BLAST's primary scientific motivations are to study the
spatial and redshift distribution and evolution of high-redshift star-forming
galaxies and to probe the earliest stages of star formation within Galactic
molecular clouds.

BLAST conducted a 250-hour flight, launching from McMurdo Station, Antarctica
on 2006 December 21, and landing on the Antarctic Plateau 2007 January 2
(BLAST06)\@.  BLAST flew at an average altitude of 38.6\,km with diurnal
variations between 37.5 and 39.6\,km.  Several extragalactic and Galactic
fields were mapped, including two large ($8.7\,\mathrm{deg}^2$) and one deep,
confusion-limited ($0.8 \,\mathrm{deg}^2$) extragalactic fields and two large,
overlapping regions (a $50 \,\mathrm{deg}^2$ deep region and a
$200\,\mathrm{deg}^2$ wide region) in the direction of Vela
\citep{devlin2009,netterfield2009}.

The primary science goals of the BLAST experiment demand an absolute flux
calibration accuracy of 5--10\% in all three BLAST pass-bands. 
In particular, a target, uncorrelated 5\% uncertainty is driven by the
extra-galactic science case to enable precise measurements of colors and
thereby constrain the bolometric luminosities and star-formation rates of
distant galaxies \citep{hughes2002}.
In order to achieve these goals, a primary calibration source for BLAST
with the following properties was required: (i) point-like and bright
enough to be detected in each band with a SNR exceeding 20$\sigma$ in a
single map; (ii) the absolute (correlated) uncertainty in the SED had to be
less than 10\%; and (iii) the uncorrelated components of the uncertainty in
the SED (i.e.\ uncertainties in the ratios of flux densities in different
BLAST bands) could be no greater than 5\%.
\defcitealias{truch2008}{T08}
In this paper, we report on the calibration and performance of BLAST06,
concentrating on the differences from the calibration procedures used in
BLAST05, discussed in \citet{truch2008} (hereafter \citetalias{truch2008}).  
Section~\ref{sec:red} outlines the basic reduction steps and characterization of BLAST06 data.
Section~\ref{sec:perf} discusses the performance of the warm optics in BLAST06.
Section~\ref{sec:point} outlines the pointing performance of BLAST06.
Section~\ref{sec:fluxcal}, the bulk of this paper,  describes in detail the 
absolute calibration (from detector Volts to Jy on the sky) derived from  
the primary flux calibrator VY\,CMa \citep{hoffmeister1931,guthnick1939}.

\section{Data Reduction} \label{sec:red}
The data reduction for BLAST is discussed in detail in \citet{pascale2008},
\citet{patanchon2008}, and \citetalias{truch2008}.  Briefly, the data from BLAST
consist of a set of 288 bolometer time streams, in voltage units, sampled at
$100\,$Hz.  Eighteen of these time streams are diagnostic channels, useful for
removing common mode noise, the remaining 270 are coupled to the telescope.
These bolometer data are first cleaned for post-flight analysis by being
de-spiked and then deconvolved to remove the effects of the
data acquisition system filters from the
timestreams.  The cleaned data are then combined with a post-flight pointing
solution \citep{pascale2008} to make maps at each wavelength.  The map-making
process takes advantage of the multiple detectors, as well as significant scan
cross-linking, to minimize striping due to instrumental drifts
\citep{patanchon2008}.

The bolometers in each array are corrected for relative gains, or flat-fielded,
so that multi-bolometer maps can be generated.  The flat-field corrections are
determined by making individual maps for each bolometer from a single
point-source calibrator, in this case, VY\,CMa (see \S~\ref{sec:vycma}).  The
bolometers are also corrected for responsivity variations over time by using
the signals from a calibration lamp in the optics box which was pulsed every 15
minutes throughout the observations.  The resulting signal is used to correct
any time-varying changes in responsivity per bolometer.  Both the time
varying changes and the variations of beams across each array are small and the
amount of variation is comparable to those detailed in \S~2 of \citetalias{truch2008}
for BLAST05.

To calculate the flux density from a point source we adopt a matched-filtering
technique similar to that used to extract point sources from several recent
extra-galactic submillimeter surveys \citep[e.g.,][]{coppin2006,scott2006} and
detailed in \S~2 of \citetalias{truch2008}.  The beam profile on the sky, or point
spread function (PSF), used for calibration and flux extraction is generated by
stacking and averaging several observations of VY\,CMa after removing the background.
The PSF is normalized as in Equation~1 of \citetalias{truch2008}; if maps
are further filtered, the PSF used for analysis of that map must also be
filtered after normalization.  See \S~\ref{sec:vycma} for discussion of the
background-subtraction technique.  

Because the BLAST filters have wide (30\%) spectral bands \citep[\S~2.5 and
Figure~2]{pascale2008}, the derived flux densities depend on the shape of the source
spectrum within the band, and thus on the temperature of the source.
A correction must be made to calculate monochromatic flux
densities from the maps.  See \S~5.1 of \citetalias{truch2008} for details of
this color-correction.  

\section{BLAST06 Warm Optics Performance} \label{sec:perf} 
\begin{figure*}
\begin{center}
\includegraphics[angle=0, width=6.4in]{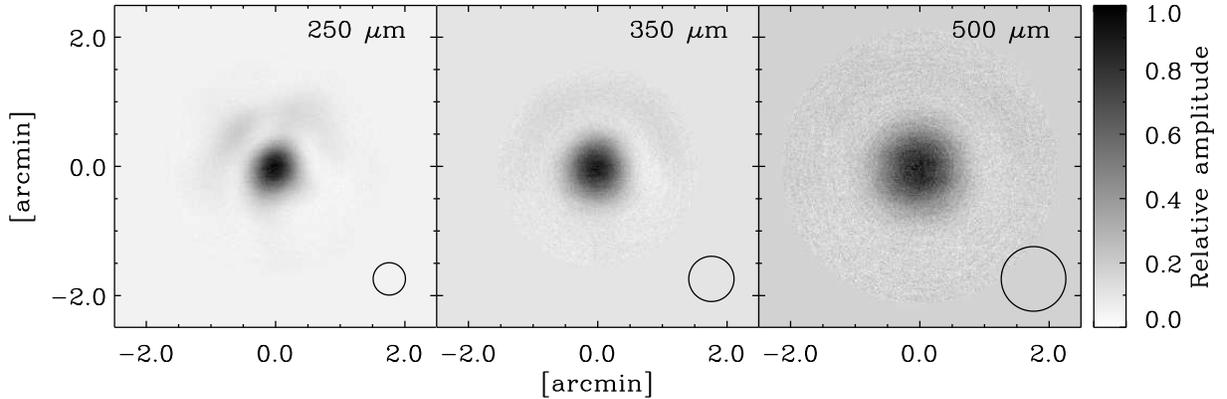}
\caption{Point Spread Functions (PSFs), provided by observations of VY\,CMa
(\S~\ref{sec:vycma}), for each of the three wavebands from BLAST06, 250, 350,
and 500\,\micron\ from left to right. These are generated by stacking several point
source maps in telescope coordinates and removing the background
(\S~\ref{sec:back}).  The small circles represent the expected diffraction limited
FWHM for each of the wavebands, 30, 42, and 60\arcsec, respectively
\citep{pascale2008}.  Fitting Gaussians to the PSFs results in FWHMs of 36, 42,
and 60\arcsec\ which contain 76\%, 92\%, and $>$\,95\% of the power,
respectively, indicating the relative power in the sidelobes, which are most
visible in the 250\,\micron\ PSF\@.  Each PSF has been set to zero outside a
radius much greater than its FWHM (1\farcm5, 1\farcm5, and 2\farcm0,
respectively) such that $>$\,98\% of the power is represented.  Cutting at
larger radii begins to include residual noise at larger scales not removed by
the background template.  
\label{psfs}
}
\end{center}
\end{figure*}

The BLAST06 warm optics (primary and secondary mirrors) performed within their
specifications.  The BLAST06 optical performance was very much improved over that of
BLAST05 through the use of a new aluminum primary mirror as well as an
in-flight focusing system \citep[\S~2.4]{pascale2008}.  
PSFs for each of the BLAST bands are shown in Figure~\ref{psfs}.  

With the pre-flight predicted Noise Equivalent Flux Density (NEFD) of 220
mJy\,${\mathrm s}^{1/2},$ we would expect 1-$\sigma$ surface brightness
fluctuations at the nominal resolutions of ${\sim}\,\rm{NEFD}/\Omega,$ or 11,
4.7, and 2.6 MJy sr$^{-1}\,{\mathrm s}^{1/2}$ at 250, 350, and 500\,\micron,
respectively where $\Omega$ is the area of the beam (FWHM$^2$).  
The measured sensitivities in each band were 8.8, 4.8, and $2.7\,{\rm
MJy\,sr^{-1}}\,{\mathrm s}^{1/2}$, respectively, which demonstrates that both
the detectors and optical efficiencies largely achieved our design goals.  The
only exception is the 250\,\micron\ beam which is slightly larger than the
diffraction limit, and has significant side-lobes.
For each PSF, 76\%, 92\%, and $>$\,95\% of the power is located within a
Gaussian fit to the PSF, which have FWHM of 36, 42, and 60\arcsec\ at 250, 350,
and 500\micron, respectively.  See Table~\ref{calib}.
See \S~3 and Figure~4 of \citet{pascale2008} for further details on the
noise and sensitivity of the BLAST detectors.  

\begin{deluxetable*}{ccccccc}
\tablecaption{Calibration and Performance Parameters for BLAST06\label{calib}}
\tablewidth{0pt}
\tablehead{
\colhead{Band} & \colhead{calib.~coeff.} & \colhead{uncertainty\tablenotemark{a}} &
\multicolumn{3}{c}{Pearson correlation matrix\tablenotemark{b}}  & \colhead{PSF FWHM}\\
\colhead{[\micron]} & \colhead{[$\times 10^{12}$ Jy V$^{-1}$]} & \colhead{[\%]} &
\colhead{250\,\micron} & \colhead{350\,\micron} & \colhead{500\,\micron} & \colhead{[arcsec]}
}
\startdata
250   &  2.73 &   9.5   &   1   &  0.83  &  0.80 & 36 \\
350   &  2.86 &   8.7   &       &  1     &  0.83 & 42 \\
500   &  1.16 &   9.2   &       &        &  1    & 60 \\
\enddata
\tablecomments{Calibration coefficients, calibration uncertainties, and Pearson
correlation matrix, showing the relationship between errors in different bands,
for BLAST06.  Note the high correlation means that the measurements involving
the ratio of BLAST brightnesses, such as spectral indices and temperature, can
be reported with much higher accuracy.  Also included are the measured FWHM of
each beam determined by fitting a Gaussian; 1-$\sigma$ errors on the FWHM are
of order 1\arcsec.}
\tablenotetext{a}{These values include the 5\% uncertainty estimated for the
  bandpass measurements.  Without them, they are 8.1\%, 7.1\%, and 7.8\% at
  250, 350, and 500\,\micron, respectively.}
\tablenotetext{b}{These values include the 5\% uncertainty correlated at 50\%
  between bands for the bandpass measurments.  Without them, they are 0.977
  between 250 and 350\,\micron, 0.917 between 250 and 500\,\micron, and 0.981
  between 350 and 500\,\micron.}
\end{deluxetable*}

\section{BLAST06 Pointing Performance} \label{sec:point}
Pointing is measured in-flight to an accuracy of $\sim$ 30\arcsec\ rms using a
combination of fine and coarse sensors, including fiber-optic gyroscopes,
optical star cameras, a differential GPS, magnetometer, and Sun sensor
\citep[\S~7]{pascale2008}.  Post-flight pointing reconstruction uses only the
gyroscopes and day-time star cameras.  The algorithm is based on a similar
multiplicative extended Kalman filter technique used by {\it WMAP}
\citep{pittelkau2001,markley2003}, modified to allow the evaluation of the
alignment parameters of the star cameras and gyroscopes
\citep[\S~11]{pascale2008}.
The offset between the star cameras and the submillimeter telescope was
measured by repeated observations of pointing calibrators throughout the
flight.  We find that the relative pointing between the star cameras and
submillimeter telescope varies as a function of telescope elevation.  We apply
an elevation-dependent correction to pitch and yaw with approximate
peak-to-peak amplitudes of 260\arcsec\ and $\sim$ 36\arcsec, respectively, over
the full 25--60\degr\ elevation range of the telescope.  

\begin{figure}
\begin{center}
\includegraphics[angle=270, width=3in]{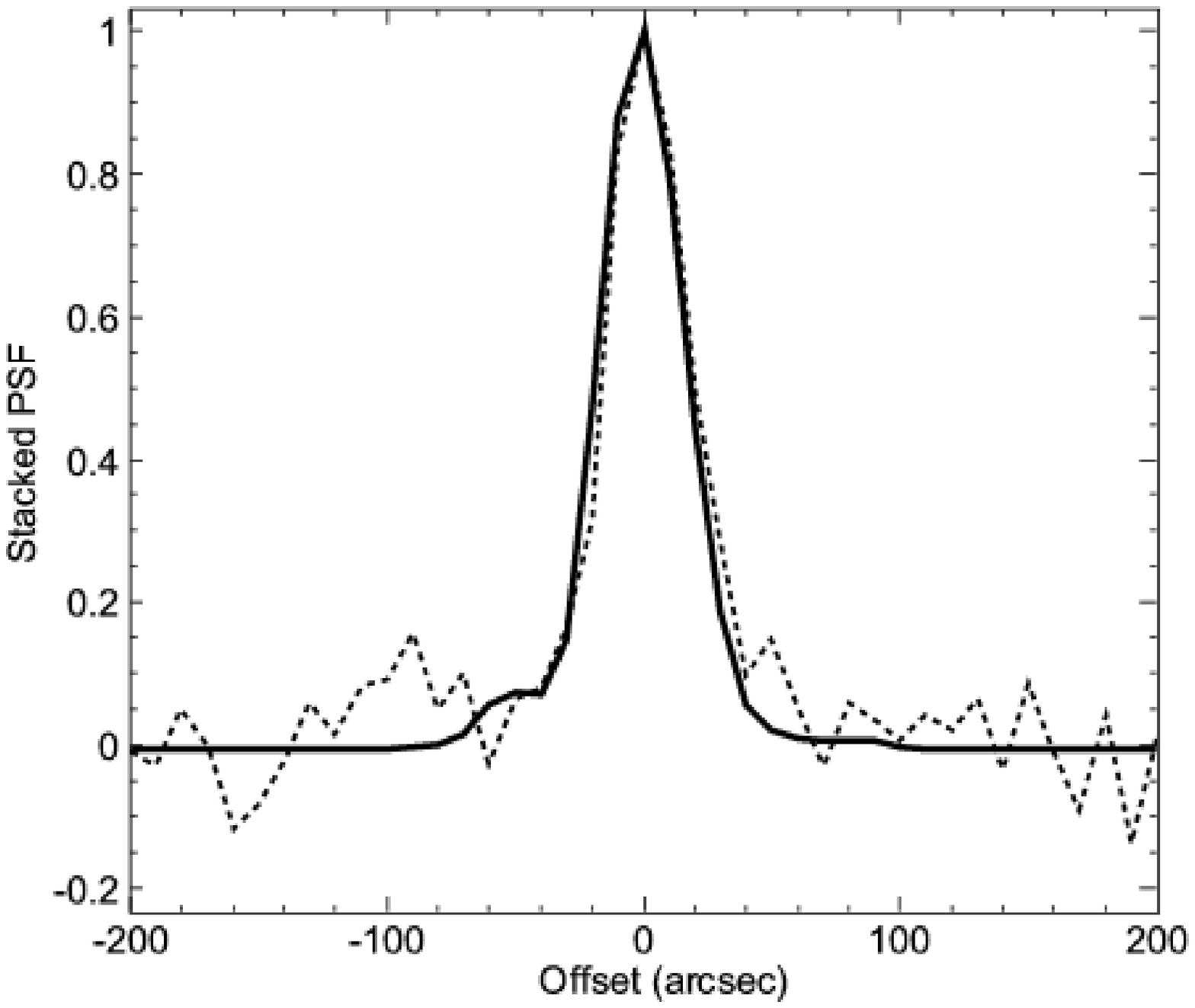}
\caption{ A cut through the peak-normalized stacked BLAST 250\,\micron\ map at the
positions of VLA 1.4\,GHz radio sources (dashed line) and through the
250\,\micron\ PSF (solid line).  We see that the stack is very well described
by the PSF, in both position and width.  We conclude that our absolute pointing
is good to 2\arcsec\ and that random pointing errors are $<$\,5\arcsec\ rms.
\label{fig:stack}
}
\end{center}
\end{figure}

Post-flight pointing accuracy is verified by a stacking analysis on one of the
extragalactic maps.  Using the deep radio Extended Chandra Deep Field---South
Very Large Array (E-CDF-S~VLA) survey at 1.4\,GHz \citep{miller2008} we stack
patches of the BLAST maps centered at the radio source coordinates, 
summing the flux density pixel by pixel (see Figure~\ref{fig:stack}).  We find that the
peak in the stacked map is located within 2\arcsec\ from the nominal position
of the catalog, more than 15 times smaller than the beam size.  Moreover,
assuming random Gaussian pointing errors, we superimpose the synthetic scaled
PSF on the stacked map and convolve it with a Gaussian profile, modeling the
broadening of the PSF due to pointing jitter.  By varying the jitter width, we
compute the $\chi^2$ of the convolved PSF over the stacked data.  In this way
we estimate the upper limit in potential random pointing errors to be $<$\,5\arcsec.  

\section{Astronomical Flux Calibration} \label{sec:fluxcal}

The primary scientific goals of the BLAST experiment demand an absolute flux
calibration accuracy of $\sim$10\% in all three BLAST pass-bands.  Achieving
this was complicated by the variability in what part of the sky was visible to
BLAST due to the unstable projected-latitude of the telescope gondola during
the flight, the restrictions on visibility due to the Sun and Moon avoidance
criteria, the orientation of Sun-shields and other baffling, and the elevation
range (25--60\degr{}) of the gondola's inner-frame \citep{pascale2008}.
Consequently, BLAST had only limited access to the calibration sources commonly
used at submillimeter and far-infrared (FIR) wavelengths.

Since the Ecliptic plane was not visible during the BLAST06 flight, no absolute
flux calibration could be determined from observations of Uranus or
Mars, for which model SEDs are known to have systematic uncertainties
$<$\,5\% at submillimeter wavelengths \citep{griffin1993,wright2007}.  The
pre-flight strategy for achieving a 10\% calibration accuracy required the
identification of alternate Galactic and extragalactic sources that could act
as primary and secondary calibrators.  The requirements included: (i)
availability throughout the flight; (ii) considered, in some cases, as
secondary calibrators for ground-based submillimeter telescopes and FIR
satellites; (iii) well-constrained SEDs in the FIR to mm-wavelength regime,
enabling accurate interpolation of the band-averaged flux densities at BLAST
wavelengths; (iv) bright ($\gg$\,1\,Jy at 500\,\micron) and compact sources (with
respect to the BLAST beam-size, i.e. $<$\,20\arcsec) that reside in regions with
minimal spatial-structure in the Galactic foregrounds or backgrounds, allowing
accurate subtraction of any extended emission.  Given these criteria we
scheduled regular observations throughout the flight of bright embedded
protostellar-sources and compact \ion{H}{2} regions, as well as the star
VY\,CMa.  For calibration purposes, we assume that the SEDs are smooth and
neglect any contribution from molecular emission lines \citepalias[See
\S\,4.1 of][]{truch2008}.  

\begin{figure}
\begin{center}
\includegraphics[width=\linewidth]{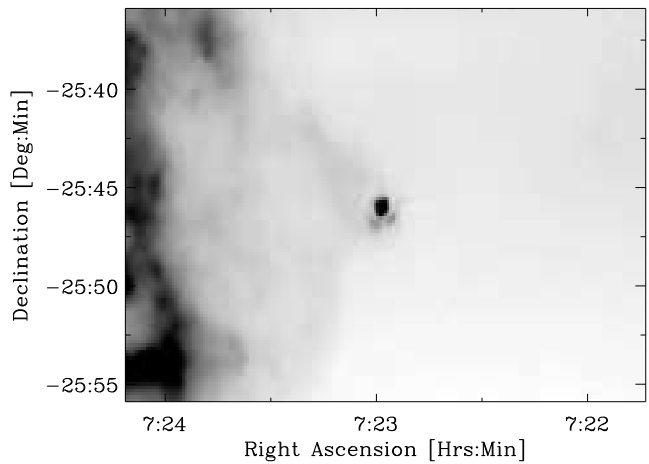}
\caption{BLAST map of VY\,CMa at 250\,\micron.  This map is in sky coordinates
and does not have the background removed.
\label{map}
}
\end{center}
\end{figure}

\begin{figure}
\begin{center}
\includegraphics[width=\linewidth]{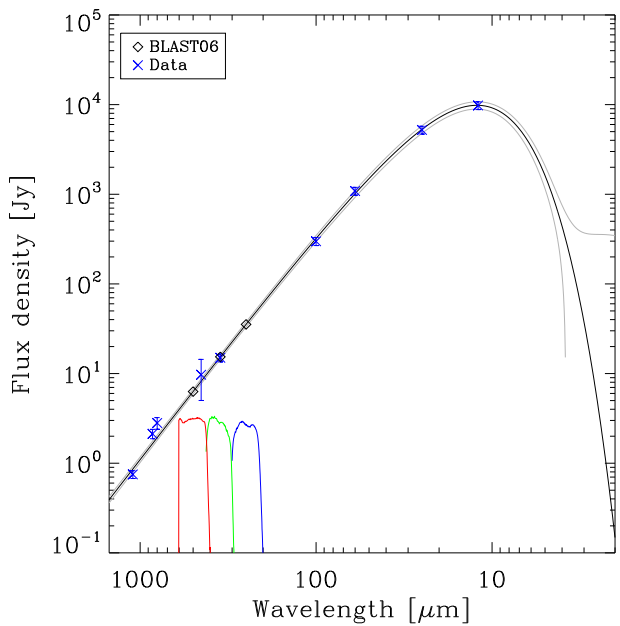}
\caption{Spectral Energy Distribution (SED) of VY\,CMa, the absolute flux
calibrator for BLAST06. The best-fit model (heavy solid-line) is constrained
using the published data (blue crosses, discussed in \S~\ref{sec:vycma},
listed in Table~\ref{fluxen}), excluding BLAST06 measurements. The grey-lines
show the 68\% confidence interval, estimated from 100 Monte-Carlo simulations,
about the best-fit model. Black diamonds indicate model predictions for BLAST06
at 250, 350, and 500\,\micron. The 1-$\sigma$ uncertainties associated with
these predictions are 8.1\%, 7.1\%, and 7.8\% in each band, respectively, which
are highly correlated.  For reference, the three BLAST passbands are shown
(normalized to an arbitrary amplitude on the plot).  
\label{sed:vycma}
}
\end{center}
\end{figure}

\subsection{VY\,CMa --- The Primary Calibrator for BLAST06} \label{sec:vycma}
We chose VY\,CMa, a red hypergiant star, as our primary calibrator.  It was
the most isolated, point-like bright object available during the flight.  It is
located in the Galactic Plane at RA $7\h 22\m 58\fs3$, Dec $-25\degr 46\arcmin
3\farcs2$ (J2000) and is one of the most intrinsically luminous stars known
\citep{humphreys2007,choi2008}.  Unfortunately, it is not entirely
point-like due to the Galactic cirrus, which significantly complicates calibration.  
Figure~\ref{map} shows the BLAST VY\,CMa map.  An
accurate PSF is required to use the calibration method as outlined above and in
\citetalias{truch2008}.  The procedure used to remove the background around VY\,CMa
is discussed below, in \S~\ref{sec:back}.

Data are collated to generate the FIR SED used to calibrate BLAST06 in a manner
similar to Arp 220 from BLAST05 \citepalias[\S~4.1]{truch2008} (see Table~\ref{fluxen} and
the SED in Figure~\ref{sed:vycma}).  These data consist of measurements from
\IRAS\ at 12, 25, 60, and 100\,\micron\ \citep{helou1988}, from SCUBA on the
JCMT at 850\,\micron\ (T.~Jenness, private communication), from SHARC-II on the CSO at
350\,\micron\ (D.~Dowell, private communication), from Bolocam on the CSO at 1.1\,mm
(J.~Aguirre, private communication), and from UKT14 on the JCMT at 450 and 
800\,\micron\ \citep{sandell1994}.
Careful attention has been placed on calculating the full correlation matrix of
the errors, as discussed in \S~4.1 of \citetalias{truch2008}.  Briefly, measurements
from the same instrument are taken to be 100\% correlated (in addition
to statistically independent photometric uncertainties), and all measurements
are assumed to have an additional fully correlated error of 5\%, since the bulk
of these instruments were all calibrated to the same Uranus SED
\citep{griffin1993}.  

\begin{deluxetable*}{ccccc}
\tablecaption{Submillimeter flux densities for VY\,CMa\label{fluxen}}
\tablewidth{0pt}
\tablehead{
\colhead{Wavelength} & \colhead{Flux Density} & \colhead{Error} & \colhead{Instrument} & \colhead{Reference}\\
\colhead{[\micron]} & \colhead{[Jy]} & \colhead{[Jy]} &
}
\startdata
12    & 9919 & 992 & \IRAS &     \citep{helou1988} \\
25    & 6651 & 665 & \IRAS &     \citep{helou1988} \\
60    & 1453 & 145 & \IRAS &     \citep{helou1988} \\
100   &  331 &  33 & \IRAS &     \citep{helou1988} \\
350   & 15.1 & 1.5 & SHARC\,II & (D.~Dowell, priv.\ comm.) \\
450   &  9.7 & 4.9 & UKT 14 &    \citep{sandell1994} \\
800   & 2.81 & 0.4 & UKT 14 &    \citep{sandell1994} \\
850   & 2.13 & 0.3 & SCUBA &     (T.~Jenness, priv.\ comm.) \\
1100  & 0.75 & 0.08 & BOLOCAM &  (J.~Aguirre, priv.\ comm.) \\
250\tablenotemark{a}   & 37.4 & 3.7 & BLAST & This paper \\
350\tablenotemark{a}   & 15.0 & 1.7 & BLAST & This paper \\
500\tablenotemark{a}   & 6.66 & 0.9 & BLAST & This paper \\
\enddata
\tablecomments{Flux densities used to generate the VY\,CMa SED shown in
Figure~\ref{sed:vycma} as discussed in the text.  Flux densities for BLAST and
from \IRAS\ have been color-corrected.  The error column shows photometric and
calibration uncertainties added in quadrature for each non-BLAST measurement;
only calibration uncertainty is shown for BLAST.
}
\tablenotetext{a}{Extracted from the fit and used to calibrate BLAST06.}
\end{deluxetable*}

We find a best-fit model with parameters 
$T = 346\pm19$\,K, $\beta = 0.545\pm0.046$, and
$S_{\mathrm{FIR}} = (3.42\pm0.38)\times10^{17}$\,W\,m$^{-2}$
(see Figure~\ref{sed:vycma}).
This results in BLAST calibration uncertainties of 8.1, 7.1 and 7.8\% at
250, 350, and 500\,\micron, respectively. These are highly correlated, with Pearson
correlation coefficients of $\rho_{250-350} = 0.977$, $\rho_{250-500} =
0.917$, and $\rho_{350-500} = 0.981$.  See Table~\ref{calib}.
The large degree of correlation in these values shows that
almost all of the uncertainty in the SED is in the absolute value. The {\em
relative} calibration uncertainty between different BLAST bands (i.e.\ 
color uncertainty), on the other hand, is actually very small, and exceeds
the requirements for the experiment. In fact, care must be taken when
fitting simple model SEDs to high-SNR measurements where the calibration
uncertainties dominate. In such cases, the BLAST data may place such
stringent constraints on SED shapes (e.g. slope and curvature through the
BLAST bands) that single-temperature thermal SEDs are not consistent with
the observations.
Section~4.5 of \citet{wiebe2009} 
discusses how to incorporate the bandpass uncertainties for high-SNR sources.

We include an additional calibration uncertainty related to our measurement
of the BLAST spectral bandpasses.  Since we have used our measured
bandpasses to calculate the incident power on our detectors given the SED
of our primary calibrator, any sources with different SEDs could have
uncertainties in their reported band-averaged flux densities arising from 
uncertainties in the BLAST bandpass measurements 
\citepalias[See Eq.~5 of][]{truch2008}.  The dominant source of uncertainty in
this measurement is our knowledge of the location of the band edges 
\citep[See Fig.~2 of][]{pascale2008}, which we believe to be no 
greater than about 5\%.  Assuming this maximum uncertainty in the measurement, 
we have calculated the effect it would have on sources with different SEDs,
ranging from ``hot'' sources ($T>$40\,K), to ``cool'' sources ($T\approx$15\,K). 
Since all of
the BLAST filters sample the Rayleigh-Jeans regime of VY\,CMa, the SEDs of
``hot'' sources resemble the primary calibrator and this uncertainty has no
effect.  However, for ``cool'' sources, the SED turns over, particularly in
the 250\,\micron\ band.  We, therefore, include a 5\% uncertainty
due to the uncertainty in the knowledge of the bandpasses.
Since the bandpasses were measured simultaneously with the same Fourier
transform spectrometer, there may be significant band-band correlations in the
bandpass measurements. We assume correlation coefficients of 50\% between all
bands.

\subsection{VY\,CMa Background Subtraction} \label{sec:back}
A complication with the use of VYCMa as a calibrator is that it is not entirely
point-like due to Galactic cirrus.  We attempt to use the cross-linking
(limited to an angle between scans of only 9\degr\ due to the extreme
southerly flight of BLAST06) of the region in order to separate the beam shape
from the diffuse background emission.  The former is coherent in telescope
coordinates, while the latter is coherent in sky coordinates.  By iterating
between the two coordinate systems, it is possible to better separate the two
components, since features in the sky are smoothed out to lower significance
levels in telescope coordinates and features in the PSF are smoothed out to lower
significance in sky coordinates.  

A zeroth-order background subtracted map is generated by extrapolating the
surrounding emission into the region of the point source.  A low-order
2-dimensional polynomial is fitted to the map with the source masked out.  

We start with the true data stream, $d,$ and an initial estimate of the PSF,
$P_0,$ taken from the zeroth-order background subtracted maps.  From there, we
scan the estimate of the point source (PSF), $P_i,$ using a sky simulator (an
inverse map-maker which takes maps and the BLAST pointing solution and
generates raw timestreams).  The iteration index is denoted by $i.$  The PSF
maps are in telescope coordinates, so scanning must also be done in telescope
coordinates.  The result is a simulated PSF-only data stream, $p_i^*.$  We then
subtract $p_i^*$ from the true data stream, $d,$ producing the background-only
data stream, $b_i = d - p_i^*$.  Then we create a point-source subtracted
background map, $B_i,$ from the background-only data stream.  This map is in
sky coordinates.  The central region of the background-only map, $B_i,$ can be
optionally smoothed, but this is only done in the first iteration.  Next, the
estimate of background, $B_i,$ is scanned with the sky simulator, producing the
simulated data stream, $b_i^*$.  Then, the background-only data stream is
subtracted from the true data stream, producing the PSF-only data stream, $p_i
= d - b_i^*$.  Finally, a new estimate of the PSF, $P_{i+1},$ is made using the
simulated data stream $p_i$.  

We performed 10 iterations of this algorithm in all three bands. We
examined the RMS in the beam difference maps between iterations, finding
that it was less than 1\% of the RMS in the total beam map for the final
iteration. In addition, the ratio of these quantities was shown to drop
monotonically, with most of the decrease from $\sim$20\% to 1\% occuring
in the first 2--5 iterations, after which it flattened out, demonstrating
that the answer had converged.

We note that the most complicated beam pattern is at 250\,\micron. However, the
final answer is least sensitive to the details of the backround subtraction
at this wavelength because the point-source contrast increases.  Since the
beam is smaller, the brightness in a beam of the diffuse background is
reduced.  Furthermore, the diffuse dust in the ISM has a temperature of
about 20\,K \citep[e.g.]{schlegel1998}, whereas VY\,CMa has a much warmer
effective SED in the BLAST bands of 200\,K, and is therefore much brighter
at 250\,\micron.

\section{Conclusions}
For the BLAST06 flight, improved optics over the BLAST05 flight yielded
improved point-source sensitivity.  BLAST06 calibration uncertainties are
approximately 10\%, as expected.  
Since we calibrate all BLAST bands to one object (VY\,CMa) the calibration
uncertainties are highly correlated; this high correlation should
be taken into account when using BLAST photometry measurements to constrain
SEDs.

Our absolute post-flight reconstructed pointing has been shown to be good to
2\arcsec\ with random pointing errors $<$\,5\arcsec.  
Together with the fact that BLAST06
uncertainties are typically much smaller than those of other experimental
measurements in the 250--500\,\micron\ waveband this should make BLAST data
useful as a cross-correlation check for future submillimeter experiments.
Future observations of VY\,CMa with the SPIRE instrument on 
\Herschel\ \citep{griffin2004} would provide a useful cross-check of the
calibration here, and the higher resolution would improve the background
subtraction.  

\acknowledgments
We acknowledge the support of NASA through grant numbers NAG5-12785,
NAG5-13301, and NNGO-6GI11G, the NSF Office of Polar Programs, the Canadian
Space Agency, the Natural Sciences and Engineering Research Council (NSERC) of
Canada, and the UK Science and Technology Facilities Council (STFC).
We thank D.~Dowell, T.~Jenness, and J.~Aguirre for their data.  
This research has been enabled by the use of WestGrid computing resources.
This research also made use of the SIMBAD database, operated at the CDS,
France, and the NASA/IPAC Extragalactic Database (NED), operated by the Jet
Propulsion Laboratory, under contract with NASA.

\bibliographystyle{apj}
\bibliography{ms}

\end{document}